# Robustness of Quantum Random Walk Search with multi-phase matching

*Hristo Tonchev[1], Petar Danev[1]*

[1] *Institute for Nuclear Research and Nuclear Energy, Bulgarian Academy of Sciences,72 Tzarigradsko Chaussée, 1784 Sofia, Bulgaria*
   Emails: htonchev@inrne.bas.bg     pdanev@inrne.bas.bg

***Abstract:*** *In our previous works, we have studied quantum random walk search algorithm on hypercube, with traversing coin constructed by using generalized Householder reflection and a phase multiplier. When the same phases are used each iteration, the algorithm is robust (stable against errors in the phases) if a certain connection between the phases in the traversing coin is preserved, otherwise small errors lead to poor algorithm performance. Here we investigate how the robustness changes if different phases are used, depending on the current iteration number. We numerically study six different examples with different phase sequences. We show that usage of a particular sequence of phases can make the algorithm more robust even if there is no preserved connection between the phases in the traversing coin.*

***Keywords:*** *Quantum Information, Quantum Algorithms, Quantum Search, Quantum walk, Generalized Householder Reflection, Logistic Regression*

1. Introduction

Quantum random walk search algorithm (QRWS) is one of the most important applications of the quantum walk. This Oracle based algorithm finds one element that satisfies certain criteria. This probabilistic algorithm was first introduced by Shenvi et al. [1] in 2003, and is quadratically faster than all known classical algorithm. Due to the possibility quantum walk to be applied on an arbitrary structure, the behavior of QRWS was studied on different structures. Some examples are: simplex [2], hypercube [3], square and cubic grids [4], tree graphs [5] and fractal structures [6]. Quantum random walk search is also used as a subroutine in other algorithms, like the quantum algorithm for graph's isomorphism [7], the algorithm for evaluating Boolean formulas [8], and the algorithm for finding a particularly shaped subgraphs in a larger graph [9].

A variety of physical systems are used in order to construct qubits. Examples of such systems are ion traps that use metastable electron levels in order to construct qubits [10], linear optical system where single photons are used as qubits [11]. On both systems given as examples, the generalized Householder reflections can be constructed efficiently [12] [13]. Any operator can be decomposed on such reflections quadratically faster [14] than decomposition by Givens rotations [15]. Household reflections can be used in order to decompose operators with arbitrary dimensions, so they can be used to construct operators in the case of registers consisting of qudits [16].

The quantum walk [17] has many modifications, allowing it to be used in various tasks. One such modification is the walk with multiple coins, that was first proposed by Brun et al. [18]. In it different quantum coins are applied during the walk depending on the walk step number. This variant of the walk has multiple applications that include quantum state preparation [19], quantum public-key secure communication [20], quantum key distribution [21], and quantum secret sharing protocols [22].

Hill function [23] [24] is logistic regression function that has been used in different fields, like systems biology, biochemistry and pharmacology in order to empirically describe non-linear processes. This function has a similar shape to the sigmoid function, but in addition has a controllable slope of the curve. In our previous work [25] we have modified this function in order to obtain Gaussian like curve, with controllable height, width and slope. This function should not be confused with the solution of the Hill equations used in quantum mechanics [26], which is the solution of a second-order differential equation for a periodic function.

The real physical systems are not perfect and there are various sources of error. The sources are highly dependent on the quantum system used in the experimental implementation. However, broadly speaking, we can divide them in two main categories – decoherence due to interaction between the system and the environment and error coming due imperfection of the setup [27]. For example, in the case of ion traps external electric [28] and magnetic [29] fields are source of errors. Examples for error sources due to the setup are laser and qudit coherence, motion and Rabi rate fluctuations and spontaneous emission. In order to be able to make quantum computations in presence of error source (both internal or external) one should try to mitigate them by improving the experimental setup, better isolate the system or make the quantum computations more fault tolerant.

The noise lowers the performance of the quantum algorithms. When the noise passes certain threshold, the quantum algorithms stops to work at all (example for Grover's algorithm can be seen in [30]). Thus, making the quantum computation more error resistant is very important for all subfields of quantum information. It is very well studied how the errors affect quantum computation (for example by adding additional noise gates [31]) and how the errors can be fixed by quantum error corrections [15].

Decoherence in the quantum walk is also widely studied. Hitting time of quantum walk is decreased in noisy environment [32]. A theoretical study of how different type of errors (phase flip, bit flip and amplitude damping channel) affect the quantum walk on line (depending on the traversing coin's symmetries) was discussed in [33]. This work also investigates the effects of such noise in NMR and atoms in optical lattice quantum computer realizations. The effects of noise and gate imperfections in quantum dots realization of quantum walk on line are examined in [34]. Continuous time quantum random walk search on different topologies with presence of noise was studied in [35]. In their work, the authors have investigated the relation between certain quantum parameters and the algorithm's robustness. Discrete quantum walk search algorithm with systematic phase errors in the traversing coin was studied in [36]. The authors used a walk coin constructed by a generalized Householder reflection and showed that such coin is very sensitive to the phase of the generalized reflection.

In two of our works dedicated to QRWS [37] and [38], we have studied modification of the quantum random walk search algorithm on hypercube with walk coin constructed by using one generalized Householder reflection (which contains a phase) and one additional phase multiplier. Such construction of the traversing coin has two phases that can be changed. If a particular dependence between those phases is maintained, the algorithm becomes more robust (probability to find solution remains high even if there are inaccuracies in the phases). Even a linear dependence can improve the algorithm substantially, however the best result is achieved when nonlinear dependence is used. We show that with modified Hill function [25], we can approximate the probability to find solution as a function of the Householder reflection's phase. In addition, extrapolations for larger coin size are also shown.

In our work [39] we have studied the robustness of different modifications of the Grover's algorithm. Each iteration consists of two reflections constructed by using generalized Householder reflection. To construct each reflection, we use two different phases depending on the current iteration number. One example is the case where the phases in the reflections differ for even and odd number of iterations. We use fits by the modified Hill function in order to compare the robustness of those modifications for different functional dependences between phases.

Here we study the robustness of different modifications of the quantum random walk search algorithm, that have different phases (both in generalized Householder reflection and the additional phase multiplier) depending on the number of the iteration. We show that similarly to the case of the Grover's algorithm, the usage of specific sequences of phases highly increases the robustness of the quantum random walk search on hypercube. The comparison of the modifications is done again by the modified Hill function.

This work is organized as follows: In Section 2, we give a brief description of the quantum random walk search algorithm on a hypercube and review its circuit. Construction of the walk coin by generalized Householder reflection and a phase multiplier is shown in Section 3. Examples of the probability to find solution for different values of both phases are also given. The robustness of the algorithm when both phases have particular functional dependence is discussed in Section 4. The modified Hill function, is shown in Section 5. Some examples of the probability to find solution fitted with the Hill function, when the phases linearly depend on a parameter, are shown in Section 6. In the next section we present our results. First, in Section 7 we describe the quantum walk search with multiple walk coins, constructed by using generalized Householder reflection and a phase multiplier. In these studies, the phases depend on the iteration number. In Section 8 and 9 we show our numerical study for six different sequences of phases for such walk coin. Those sequences are compared among themselves and with the case when just one walk coin is used each iteration. Finally, we briefly summarize our work in the Conclusion.

## 2. Quantum Random Walk Search Algorithm

Quantum random walk search algorithm is quantum algorithm that finds a node that satisfies certain criteria in unordered database structured as a graph. This algorithm can be used on a graph with arbitrary topology and is quadratically faster than the fastest classical search algorithm. This algorithm is probabilistic and the probability to find solution depends on the searched structure. One of the most important and studied quantum walks on a structure is the one on a hypercube, because all binary numbers can be depicted as points on the hypercube.

Quantum circuit of QRWS algorithm is shown on Fig. 1:

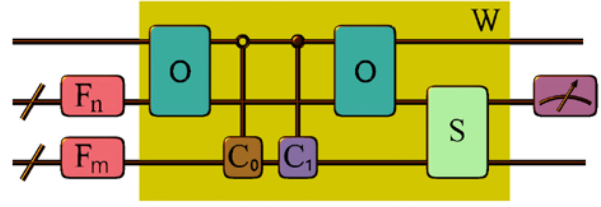

Fig. 1. *The quantum circuit of QRWS algorithm. Each search iteration W consists of an Oracle O, a Shift operator S, traversing and marking coins $C_0$ and $C_1$. The initial state of the algorithm's node and coin registers are prepared by using discrete Fourier transform operator with corresponding dimensions $F_n$ and $F_m$. At the end of the algorithm, the node register is measured.*

The algorithm has three registers. The control register is the first one, it uses only one qubit, so its dimension is 2. The third register is called edge register and corresponds to the number of edges that originates of a single node. In the case of a hypercube its dimension is equal to the dimension of the hypercube m. The second register corresponds to the number of nodes of the hypercube, its dimension can be calculated as $n = 2^m$. The size of the register will be denoted as a subscript of the ket vector. The state of the whole register will be denoted as $|\Psi_i\rangle$ where the subscript corresponds to the number of the iteration.

Initially the all registers are in the state zero:

$$|\Psi_0\rangle = |0\rangle_2 \otimes |0\rangle_n \otimes |0\rangle_m = |0,0,0\rangle_{2\times m\times n} \qquad (1)$$

Discrete Fourier transform (DFT) is applied on the node and coin registers (operators $F_n$ and $F_m$) in order to put them in equal weight superposition. The state of the control register remains unchanged.

Next, the QRWS iteration should be applied k times (the value of k will be given later in this chapter) on the whole register. The iteration consists of the following sequence of operators:

1) First an oracle ($O$) is applied to the control and node registers, entangling them. Oracle should be able to recognize if one (or more) of the nodes is a solution and mark it by changing the state of the control register. If there are $\lambda$ solutions, namely $\{h_1, h_2, \ldots, h_\lambda\}$, the action of the part containing the oracle can be written as:

$$\mathbb{O}|q,i,j\rangle_{2\times m\times n} = (O\otimes I_m)|q,i,j\rangle_{2\times m\times n} \qquad (2)$$
$$= \begin{cases} |(q+1) \bmod 2, i, j\rangle_{2\times m\times n} & j \in \{h_1, h_2, \ldots, h_\lambda\} \\ |q,i,j\rangle_{2\times m\times n} & otherwise \end{cases}$$

2) The traversing coin $C_0$ is applied on all states that are not solutions. The coin operator changes the state of the coin register and in this way changes the probability to go at each direction. For traversing coin can be used any unitary operator with dimension equal to the one of the coin register. Which one will be the best, depends on the topology and dimension of the walk structure. In the case of a hypercube, Grover coin $G$ gives the best results.

$$C_0 = G = \hat{I}_m - 2|\chi\rangle\langle\chi| \qquad (3)$$

$$\mathbb{C}_0 = \begin{pmatrix} \hat{I}_n \otimes C_0 & \hat{0}_{m \times n} \\ \hat{0}_{m \times n} & \hat{I}_{m \times n} \end{pmatrix} = Diag(\hat{I}_n \otimes C_0, \hat{I}_{m \times n}) \qquad (4)$$

Here $\hat{0}_{m \times n}$ are matrices with dimension $m \times n$ filled with zeros, $\hat{I}_n$ is the identity matrix with size $n$ and $|\chi\rangle$ is equal weight superposition, $Diag(\hat{I}_n \otimes C_0, \hat{I}_{m \times n})$ is a block diagonal matrix with the following blocks $\hat{I}_n \otimes C_0$ and $\hat{I}_{m \times n}$.

3) The marking coin $C_1$ is applied on the states that are solutions. The best marking coin depends on the chosen topology of the structure and of the chosen traversing coin. In the case of a hypercube and Householder traversing coin, the best marking coin is:

$$C_1 = -\hat{I}_m \qquad (5)$$

$$\mathbb{C}_1 = \begin{pmatrix} \hat{I}_{m \times n} & \hat{0}_{m \times n} \\ \hat{0}_{m \times n} & \hat{I}_n \otimes C_1 \end{pmatrix} = Diag(\hat{I}_{m \times n}, \hat{I}_n \otimes C_1) \qquad (6)$$

4) The oracle is applied a second time on the control and node registers in order to disentangle them.
5) At the end of the iteration a Shift operator is applied. It defines the walk structure (by determining the nodes connected by an edge) and executes the quantum walk itself depending of the state of the edge register.

$$\mathbb{S}|0, j, k\rangle = (\hat{I}_2 \otimes S)|0, j, k\rangle = |0, j, g(k, j)\rangle \qquad (7)$$

Together, the iteration $W$ of the quantum algorithm gives:

$$W = \mathbb{S} \cdot \mathbb{O} \cdot \mathbb{C}_1 \cdot \mathbb{C}_0 \cdot \mathbb{O} \qquad (8)$$

The state of the whole register at $(i+1)$-th iteration can be obtained from the i-th iteration by using the following recursive formula:

$$|\Psi_{i+1}\rangle = W|\Psi_i\rangle \qquad (9)$$

When there is only one solution, the number of iterations needed $k$ can be calculated by the following formula:

$$k_{iter} = \left\lceil \frac{\pi}{2}\sqrt{n/2} \right\rceil = \left\lceil \frac{\pi}{2}\sqrt{2^{m-1}} \right\rceil \qquad (10)$$

Where $\lceil m \rceil$ denotes the rounded-up $m$.

After the required number of iterations, the algorithm ends with measurement of the node register. The probability to find solution is approximately:

$$p = 1/2 - \mathcal{O}(1/2^m) \qquad (11)$$

If the final state do not give the right solution, the algorithm can be repeated.

### 3. Traversing coins leading to high robustness

In our previous work [37] we have studied the probability to find solution of QRWS algorithm with walk coin constructed by generalized Householder reflection with a phase multiplier:

$$C_0(\phi, \zeta, m) = e^{i\zeta}\left(\hat{I}_m - (1 - e^{i\phi})|\chi\rangle\langle\chi|\right) \quad (12)$$

The operators used in the construction of such coins can be implemented efficiently in some physical systems such as ion traps and photonic quantum computers.

When $\phi = \zeta = \pi$ we obtain the Grover coin. However, defects in the laser or the optical system can lead to inaccuracies in the phases of the coin. It is therefore important to investigate how robust (stable against phase inaccuracies) the implementation of such a coin is. The probability to find solution for such coin with arbitrary angles is:

$$p(\phi, \zeta, m) = \frac{1}{2} - f(\phi, \zeta) \times \mathcal{O}\left(\frac{1}{2^m}\right) \quad (13)$$

Here we will also study operators similar to the ones defined by Eq. (12), where we flip the sign of one of the phases:

$$C_0(-\phi, \zeta, m) = e^{i\zeta}\left(\hat{I}_m - (1 - e^{-i\phi})|\chi\rangle\langle\chi|\right) \quad (14)$$

$$C_0(\phi, -\zeta, m) = e^{-i\zeta}\left(\hat{I}_m - (1 - e^{i\phi})|\chi\rangle\langle\chi|\right) \quad (15)$$

$$C_0(-\phi, -\zeta, m) = e^{-i\zeta}\left(\hat{I}_m - (1 - e^{-i\phi})|\chi\rangle\langle\chi|\right) \quad (16)$$

The probability to find solution of the unmodified QRWS as a function of the phases for coin sizes 4 (on the first row), 6 (on the second row) and 8 (on the third row) are shown on *Fig. 2*. The left pictures correspond to the coins constructed by using equation (12) and (16). Similarly, the right pictures - to coins made according to (14) and (15). The legend in the center shows the probability to find solution as a function of the color.

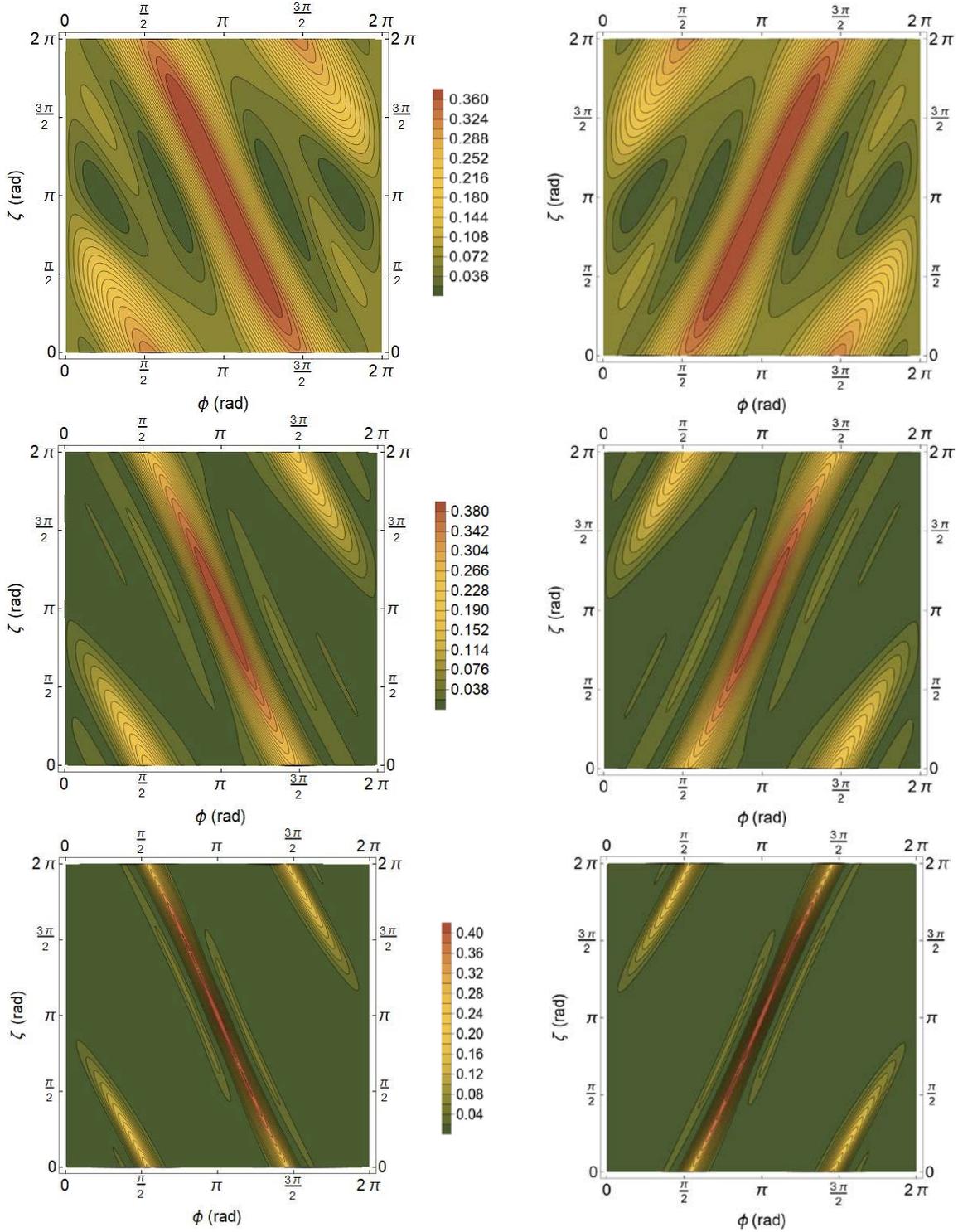

*Fig. 2. Probability to find a solution as a function of the generalized Householder phase φ and the additional phase multiplier ζ. The first row shows the results for coin size 4 and the second - for coin size 6. The left pictures correspond to coin constructed by using equations (12) and (16) and the right - by using equations (14) and (15). The legend in the center shows the correspondence between the probability to find solution and the colors on the picture.*

From the pictures on *Fig. 2*, the following observations can be made: in both cases (when the signs of the phases are the same, and when they differ) there is a stripe that has high probability to find solution and the width of the stripe depends on the coin size. Those stripes depend on the sign of the phases used in the coin construction. The stripe when the traversing coin is constructed according to Eq. (12) coincides with the stripe when Eq. (16) is used instead. Similarly, the stripe is the same for Eq. (14) and Eq. (15).

If during the QRWS both phases stay on this stripe even if there are errors in them, we can have larger inaccuracies without affecting the probability to find solution. This stripe can be approximated with a functional dependence of between the angles $\zeta$ and $\phi$:

$$p(\phi, \zeta, m) = p(\phi, \zeta(\phi), m = const) = p(\phi) \quad (17)$$

In general, such functional dependence is not a bijection, and it cannot be used when the angle between the line $\zeta(\phi)$ and the axis $\phi$ is close to $\pi/2$. That is why we define the parameters $\omega$ and $\Theta$ and both angles $\phi$ and $\zeta$ become a function of those parameters (so $\phi(\omega, \Theta)$ and $\zeta(\omega, \Theta)$). The equation of a line that passes through the point with coordinates $\{\pi, \pi\}$ and makes angle with the axis $\phi$ equal to $\Theta$:

$$\left| \begin{array}{l} \phi = \pi + \omega \cos(\Theta) \\ \zeta = \pi + \omega \sin(\Theta) \end{array} \right. \quad \& \quad \left| \begin{array}{l} \phi \in [0, 2\pi] \\ \zeta \in [0, 2\pi] \end{array} \right. \quad (18)$$

In order to achieve the conditions $\phi \in [0, 2\pi]$, $\zeta \in [0, 2\pi]$ we define:

$$\left| \begin{array}{l} \Theta \in [0, \pi] \\ \omega \in \left[ -min\left(\frac{\pi}{\sin(\Theta)}, \frac{\pi}{\cos(\Theta)}\right), +min\left(\frac{\pi}{\sin(\Theta)}, \frac{\pi}{\cos(\Theta)}\right) \right] \end{array} \right. \quad (19)$$

where $min(a, b)$ is the smallest number between $a$ and $b$.

In this form, by using the angle $\Theta$ and amplitude $\omega$, the whole plain could be covered and the fixed functional dependence between phases used in our previous works is equivalent to a fixed angle $\Theta$. Our previous parametrization required usage of different basis when the angle between $\zeta(\phi)$ and the axis $\phi$ is large (even the line $\zeta(\phi) = \pi$ is undefined). When the parametrization with $\omega$ and $\Theta$ is used, only the point $\{\pi, \pi\}$ is poorly defined.

The same stripe in the new basis can be written as functional dependence of each of the angles $\zeta$ and $\phi$ from the parameter $\omega$:

$$p(\phi, \zeta, m) = p(\phi(\omega, \Theta = const), \zeta(\omega, \Theta = const), m = const) = p(\omega) \quad (20)$$

We can introduce different functional dependences between the phases $\{\phi(\omega), \zeta(\omega)\}$ in order to best fit this stripe (such fit will give maximal stability against errors). In order to be able to compare the fits we define the robustness $\varepsilon$ as:

$$p\big(\omega \in (\omega_{max} - \varepsilon^-, \omega_{max} + \varepsilon^+)\big) \cong p_{max} \equiv p(\omega_{max}) \quad (21)$$

Here, the maximal probability to find solution for a particular functional dependence $\{\phi(\omega), \zeta(\omega)\}$ is denoted by $p_{max}$ and $\omega_{max}$ is the value of $\omega$ where $p_{max}$ is achieved. The interval $\Delta = (\omega_{max} - \varepsilon, \omega_{max} + \varepsilon)$ is symmetric about the point $\omega_{max}$, so we can simplify the above formula by using $\varepsilon^- = \varepsilon^+ = \varepsilon$.

$$p(\omega \in (\omega_{max} - \varepsilon, \omega_{max} + \varepsilon)) \geq \Omega p_{max} \qquad (22)$$

were $\Omega$ is a number close but less than one.

In the next section we will compare the robustness with different functional dependences that were first introduced in [37].

### 4. Robustness with selected functional dependences

In our previous works, the traversing coins was defined by Eq. (12) and in this case, we have calculated the robustness when a particular functional dependence between phases is present $\zeta(\phi)$. Of particular interest to us was the probability to find solution when a linear dependence is fulfilled (and give high robustness):

$$\zeta = -2\phi + 3\pi \qquad (23)$$

And the perpendicular line to the one given by Eq. (23) leads to low robustness:

$$\zeta = \phi/2 + \pi/2 \qquad (24)$$

They are all linear functional dependences and are written in the following form:

$$\zeta = C_1 \phi + C_2 \qquad \& \qquad \{\pi, \pi\} \in \{\zeta, \phi\} \qquad (25)$$

Equations (23) and (24) still hold if the traversing coins are defined according to (16).

All those relations between angles can be obtained in the new basis. For example, when $\Theta = 233\pi/360$, we approximately obtain highly robust quantum walk search, because it is equivalent to a coin given by Eq. (12) with functional dependence Eq. (23). Similarly, if $\Theta = 53\pi/360$ we obtain low robustness very close to the one we obtain with functional dependence (24) in case of coin given by Eq. (12).

We will note that in the case of walk coins constructed according to equations (14) and (15), the best and worst lines must be mirrored with respect to the $\zeta = \pi$ axis. The new relations between the phases that will lead to highest and lowest robustness are:

$$\zeta = 2\phi - \pi \qquad (26)$$

$$\zeta = -\phi/2 - 3\pi/2 \qquad (27)$$

In case of Eq. (26) the angle in the new basis is approximately $\Theta = 127\pi/360$, and in case of Eq. (27) the angle in the new basis is approximately $\Theta = 307\pi/360$ correspondingly. It is also important to note that the maximal robustness is the same no matter if the walk coin is constructed according to Eq. (12), (14), (15) or (16). Similarly, this is also true for the minimal robustness – it is the same regardless of the equation used to construct the traversing coin.

On *Fig. 4* in Section 6 are shown examples for the probability to find solution depending on the parameter $\omega$ when the value of $\Theta$ corresponds to Eq. (26) (shown in red color) or Eq. (27) (shown in green color) for different coin sizes.

## 5. Modified Hill Function

Hill function is semi-empirical nonlinear function that is used for description of nonlinear processes in various fields including pharmacology, biochemistry and systems biology. In our previous work [25] we modified it in order to be used to describe functions having Gaussian or rectangular shape. The formula for the modified Hill function is:

$$W(\varphi, b, k, n, c) = \frac{bk^n}{|\varphi - c|^n + k^n} \quad (28)$$

Each parameter in the formula corresponds to different properties of the curve. Parameter b controls its height (as is shown on *Fig. 3* top left), parameter k controls the width of the bell (as can be seen on *Fig. 3* top right) and the parameter n controls both curvature and the slope of the curve (see *Fig. 3* bottom left). The maximal height of the bell is achieved at the point $\varphi = c$ (see *Fig. 3* bottom right). We add one addition restriction, that the parameters b, k and n should be positive and $n > k$. The reasons behind those conditions are described in detail in [25].

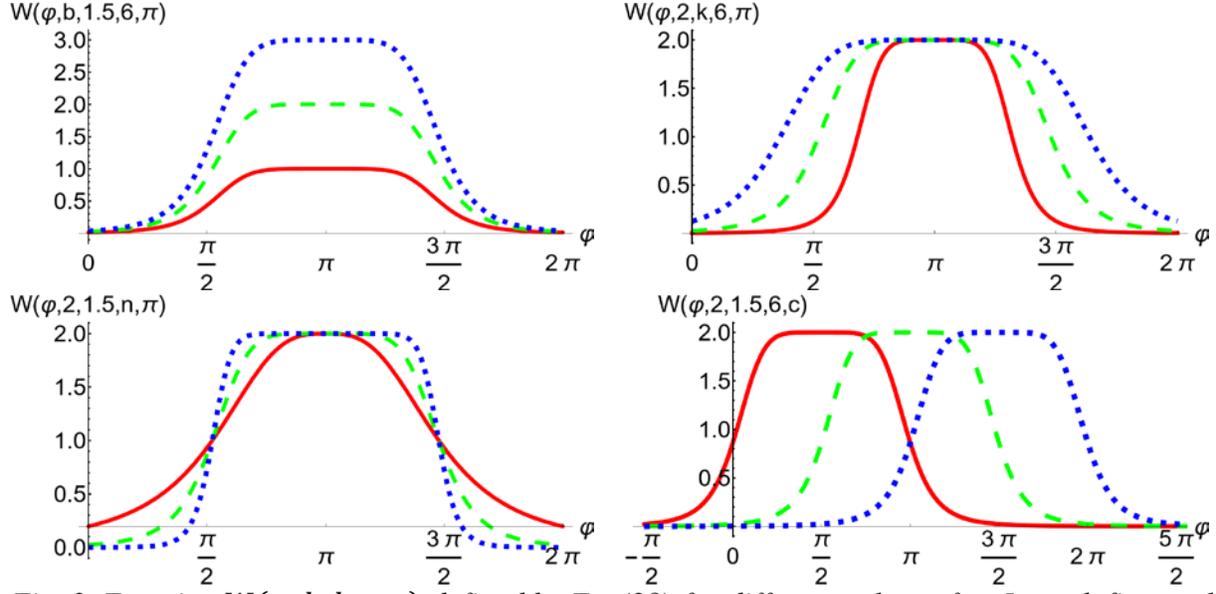

*Fig. 3. Function $W(\varphi, b, k, n, c)$ defined by Eq. (28) for different values of $\varphi$. In each figure all parameters (b, k, n, and c) are the same except one. The parameter values that are the same are written on the figure as parameters of $W(\varphi, ...)$. On the top left picture parameter b has the following values $b = 1$, $b = 2$ and $b = 3$ and the corresponding curves are shown with solid red, dashing green and dotted blue line. The top right figure shows plots with three different values of parameter k, namely $k = 1$ (solid red), $k = 1.5$ (dashing green) and $k = 2$ (dotted blue). The lines on the bottom left picture are for the following values of n: $n = 1$ (solid red), $n = 6$ (dashing green) and $n = 12$ (dotted blue). The bottom right picture shows lines with the following values of the parameter c, namely $c = \pi/2$ (solid red), $c = \pi$ (dashing green) and $c = 3\pi/2$ (dotted blue).*

The modified Hill function is very suitable for fitting and analyzing more complex functions, because each of its parameters controls one of the characteristics of the curve (central point, height, slope, and so on). Fit's parameters also give us quantitative characteristics that can be used in order to compare different bell like curves.

We use standard deviation in order to estimate how good is the fit by the Hill function:

$$\sigma = \sqrt{\sum_{j=1}^{N_p} \frac{\left(W_j(\phi_j, b, k, n, c) - p_j\left(\phi(\omega_j), \zeta(\omega_j)\right)\right)^2}{N_p - q}} \quad (29)$$

Where $q$ is the number of the parameters used in the fit, $N_p$ is the number of the points used in the fit, $p_j$ is the probability to find solutions at angle $\varphi_j$ (were $\omega_j - \omega_{j-1} = const$). This fit is made for fixed functional dependence between the phases $\{\phi(\omega_j), \zeta(\omega_j)\}$ and $W_j(\phi_j, b, k, n, c)$ is the fit by the modified Hill function.

## 6. Example of a Hill function fit for the probability in case of single-phase matching

In this section we will demonstrate how the fits with modified Hill function can be used in order to compare the robustness for the different functional dependences between the phases and how they could be used in order to make a prediction for larger coin size. Even more, we will use this method in order to compare the robustness of QRWS modifications that use the same phases at each iteration with the robustness of multi-phase QRWS.

On *Fig. 4* are shown the probabilities to find solution depending on the value of the parameter $\omega$ for walk coin defined by Eq. (12) with functional dependence given by Eq. (18) with two values of $\Theta$, namely $127\pi/360$ (represented by the green dotted line) and $233\pi/360$ (represented by the red dotted line). Their fits with the modified Hill function are shown with golden and teal dashed lines respectively. The left, central and right rows show results in case of coin sizes 4, 6 and 8. Their corresponding fit parameters are given in *Table 1*.

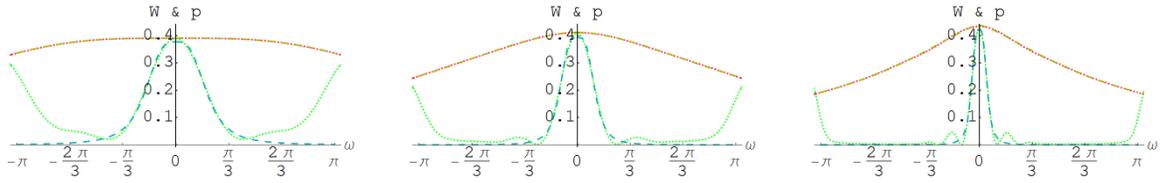

*Fig. 4. Probability to find solution as a function of $\omega$ in case of $\Theta = 127\pi/360$ (shown with doted green line) and $\Theta = 233\pi/360$ (shown with doted red line). Their corresponding Hill fits are presented with dashed golden and teal lines. The left, central and right pictures correspond to coin sizes 4, 6 and 8.*

The probability to find solution as a function of $\omega$ for angles $\Theta = 127\pi/360$ and $\Theta = 233\pi/360$ in case of coin sizes 4,6 and 8 is fitted with the modified Hill function. The fit parameters are given in *Table 1*. The coin size is given on the second column and angle $\Theta$ on the third. The parameters of the Hill fit b, k and n are shown on the fourth, fifth and sixth columns respectively. The standard deviation of the fit is presented in the seventh column.

| № | Coin Size | $\Theta$ | b | k | n | $\sigma$ |
|---|---|---|---|---|---|---|
| 1 | 4 | $127\pi/360$ | 0.377871 | 0.61954 | 3.26522 | 0.00840781 |
| 2 |   | $233\pi/360$ | 0.391042 | 5.52073 | 3.22841 | 0.000522782 |
| 3 | 6 | $127\pi/360$ | 0.393237 | 0.351782 | 3.67782 | 0.0136556 |
| 4 |   | $233\pi/360$ | 0.411837 | 4.04468 | 1.80806 | 0.00100045 |
| 5 | 8 | $127\pi/360$ | 0.421285 | 0.158438 | 3.3686 | 0.0181447 |
| 6 |   | $233\pi/360$ | 0.438595 | 2.64616 | 1.40927 | 0.00154029 |

*Table 1. Modified Hill's function fit of $P(\omega)$. The coin size is shown on the second column, $\Theta$ on the third, the next three columns show the Hill fit parameters b, k and n and standard deviation is given in the last column.*

Similarly, if we make fits of $P(\omega)$ for more coin sizes, we can use them in order to extrapolate the dependence of the Hill function parameters on the coin size (b(m), k(m) and n(m)). In this way, we can make extrapolation of the probability to find solution for $W(\Theta = 127\pi/360, m)$ and

W($\Theta = 233\pi/360$,m) for even larger coin sizes. Fits of the parameters b, k and n for coin sizes between 4 and 9 are shown on *Fig. 5*. The dashed line corresponds to parameter in case of angle $\Theta = 127\pi/360$ (lowest robustness) and the solid line - in case of $\Theta = 233\pi/360$ (highest robustness). The left, central and right pictures show the results for parameters b, k and n respectively.

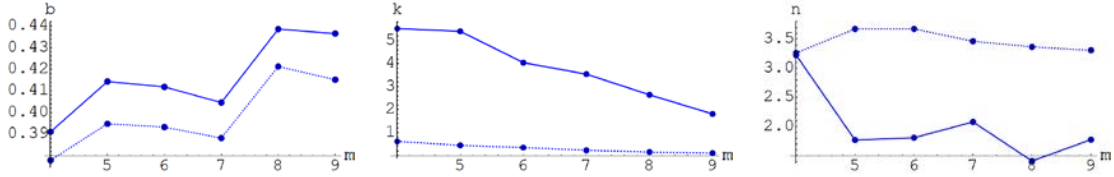

*Fig. 5. Dependence of the parameters b (shown on left), k (shown on centre) and n (shown on right) on coin size. The solid and dashed lines represent the results for angles $\Theta = 233\pi/360$ and $\Theta = 127\pi/360$ respectively. Dots show the numerical results for points with $m = $ 4,5,6,7,8 and 9.*

The Parameter n corresponds to the slope of the curve. It can be seen that the value of n for the $\Theta = 127\pi/360$ is around 3.25 for small coin sizes and slowly decreases with increasing the coin size. The value of n for $\Theta = 233\pi/360$, have high value for $m = 4$, but it decreases a lot for m=5 and overall decreases with increasing the coin size.

Parameter b corresponds to the average height of the plateau. The maximal probability of both $\Theta = 127\pi/360$ and $\Theta = 233\pi/360$ is achieved at the same point $\{\varphi(\omega), \zeta(\omega)\} = \{\pi, \pi\}$, however for $\Theta = 127\pi/360$, the average probability is smaller. This can be explained by the higher slope for the case of $\Theta = 127\pi/360$, where the averaging of the Hill fit is smaller than the one for $\Theta = 233\pi/360$.

In our investigations of robustness, the most important parameter is k. It can be seen that the robustness for angle $\Theta = 127\pi/360$ is always smaller than the one for $\Theta = 233\pi/360$ and they both decrease with increase of the coin size. The decrease is faster in case of $233\pi/360$.

Here we will use the Hill function in order to compare the parameter $k(m)$ of different modifications and angles $\Theta$. To make the comparison easier, we will fit the Hill function parameter $k(m)$ for each modification studied here by using the function:

$$k_{PM}^{Best\ or\ Worst}(m) = k_1 e^{-mk_2} + k_3 \qquad (30)$$

On *Fig. 6* we show the corresponding fits of the parameter $k(m)$ in case of $\Theta = 127\pi/360$ (right picture) and $\Theta = 233\pi/360$ (left picture). Numerical results obtained by the Hill fit are shown with teal dots and the corresponding fit of $k(m)$ by Eq. (30) are presented by a dashed line. The parameters of the fit $k_1$, $k_2$ and $k_3$ are shown in *Table 2*.

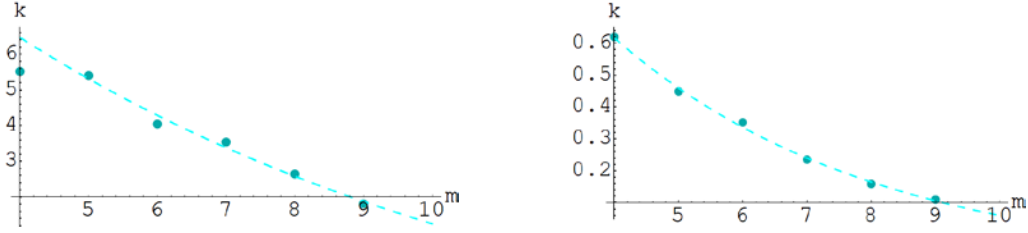

*Fig. 6. Fit of Hill fit's parameter k for coin sizes between 4 and 9. The teal points correspond to the values of k obtained from a Hill fit of the probability to find solution depending on ω. The cyan dashed line is the fit of $k(m)$ obtained by using Eq. (30). The left picture corresponds to $\theta = 233\pi/360$ and the right one - to $\theta = 127\pi/360$*

It is important to note that when fitting the parameter k for $\theta = 233\pi/360$, we exclude the point in case $k = 4$ due to different behaviour of k in the interval between 2 and 5 (for more information see [25]).

In *Table 2* are shown the results from the fit of the parameter k for phase matching and two different angles $\theta$. Fit parameters are shown in fourth, fifth and sixth columns. The standard deviation used to evaluate the goodness of the fit is shown in 7-th column.

| № | $\theta$ | $k_1$ | $k_2$ | $k_3$ | $\sigma$ |
|---|---|---|---|---|---|
| 1 | $233\pi/360$ | 16.3564 | 0.123172 | -3.52146 | 0.22472 |
| 2 | $127\pi/360$ | 1.93362 | 0.246102 | -0.105428 | 0.0121064 |

*Table 2. Fit of the parameter k for phase matching modification by Eq. (30), for angles $\theta$ shown in second column. Parameters of the fit are given in the third, fourth and fifth columns. The corresponding standard deviation is shown in last column.*

The value of $\sigma$ shows that the precision in case of $\theta = 127\pi/360$ is much higher than in case of $\theta = 233\pi/360$. That means that the prognose in case of $m \to \infty$ (that is approximately the value of the parameter $k_3$) is much better for $\theta = 127\pi/360$.

We will use the same function (Eq. (31)) in our studies in the next chapters for the following reasons: First it fits well the points. The second reason is that the parameter $k_3$, gives us the evaluations of parameter k for large coin size. And the third reason is that by using the same function for all fits allows us to make comparison easily.

## 7. Quantum random walk search with multiple walk coins

QRWS algorithm is most stable, when ω is chosen as $\omega = 0$. However, due to experimental limitations, such precise control of the phases is not always possible. The main goal of the section is to compare how the probability to find solution changes when there are inaccuracies in $\omega = 0 \pm \varepsilon_\omega$ for arbitrary $\theta$ (this correspond to the phases $\phi = \pi \pm \varepsilon_\phi$ and $\zeta = \pi \pm \varepsilon_\zeta$). We will show that we can increase the robustness by using multi-phase matching as we have already proposed for the Grover's algorithm.

A multi-phase matching for Grover's algorithm was first introduced by Toyama et. al. in their work [40]. The authors used different phases at even and odd iterations of the Grover's algorithm as the operators in the iteration are constructed by using two generalized Householder reflections. Their main goal was to show that such phase matching modification of the Grover's algorithm is more stable against changing the ratio between the register size and the number of solutions. In our work [39], we have shown numerically that by using selected phase matching modifications, we can make Grover's algorithm more stable against inaccuracies in its phases. We showed that when a particular construction is used, the algorithm remains stable against inaccuracies even for an arbitrary linear functional dependence between the phases.

Walk and traversing coins play similar role to the reflections in Grover's algorithm. In the construction of the traversing coin participate two phases and in the marking coin there is one phase. In [41] we have shown that instead of a phase multiplier in the marking coin, a change of the walk coin can be used to achieve the same interaction operator. Another important property that should be considered comes from the topology of the Hypercube [3] [42]. The quantum random walk search on hypercube, is divided on two different quantum walks one on the even iterations and second on the odd iterations. We will show here that we can increase the robustness by using quantum walk with multiple traversing coins. This means that different sequence of walk coins should be used on every two iterations.

In quantum walk search with multiple coins, the walk iteration operator depends on the current iteration. The state of the whole register at $(j + 1)$-th iteration can be obtained by using the state in j-th one and the corresponding walk coin operator:

$$|\Psi_{j+1}\rangle_N = \prod_{j=1}^{k} W_j(\phi_j(\omega,\Theta), \zeta_j(\omega,\Theta)) |\Psi_j\rangle_N \quad (32)$$

At each iteration only the traversing coin differs, the remaining operators are the same:

$$W_j(\phi_j, \zeta_j) = \mathbb{S} \cdot \mathbb{O} \cdot \mathbb{C}_1 \cdot \mathbb{C}_0(\phi_j(\omega,\Theta), \zeta_j(\omega,\Theta)) \cdot \mathbb{O} \quad (33)$$

The walk coins differ from each other by the value of the parameter $\omega$, used to obtain their phases:

$$C_0(\phi_j, \zeta_j, m) = e^{i\phi_j(\omega,\Theta)}(\hat{I}_m - (1 - e^{i\zeta_j(\omega,\Theta)})|\chi\rangle\langle\chi|) \quad (34)$$

$$\mathbb{C}_0(\phi_j(\omega,\Theta), \omega_j(\omega,\Theta)) = \begin{pmatrix} \hat{I}_n \otimes C_0(\phi_j(\omega,\Theta), \zeta_j(\omega,\Theta)) & \hat{0}_{m\times n} \\ \hat{0}_{m\times n} & \hat{I}_{m\times n} \end{pmatrix} \quad (35)$$

where the phases $\phi_j$ and $\zeta_j$ depend on: the current iteration, the parameter $\omega$ and the model used:

$$\phi_j = f(\omega, \Theta = const, j) \qquad \zeta_j = f(\omega, \Theta = const, j) \quad (36)$$

In the next two sections, we investigate the robustness against phase errors of six modifications that have the sign of the phases that depend on the current iteration. The phases will be written in the form $\{\phi_j, \zeta_j\} = \{\phi(\omega, j), \zeta(\omega, j)\}$ for simplicity.

## 8. Sequences with alternate sign flipping

The sequences with alternating sign that will be studied in this section are the following:

$$\{\phi_j, \zeta_j\} = \{\phi, (-1)^{[j/2]}\zeta\} \quad denoted\ as\ A1 \tag{37}$$

$$\{\phi_j, \zeta_j\} = \{(-1)^{[j/2]}\phi, \zeta\} \quad denoted\ as\ A2 \tag{38}$$

$$\{\phi_j, \zeta_j\} = \{(-1)^{[j/2]}\phi, (-1)^{[j/2]}\zeta\} \quad denoted\ as\ A3 \tag{39}$$

where $[j/2]$ means that $j/2$ is rounded up.

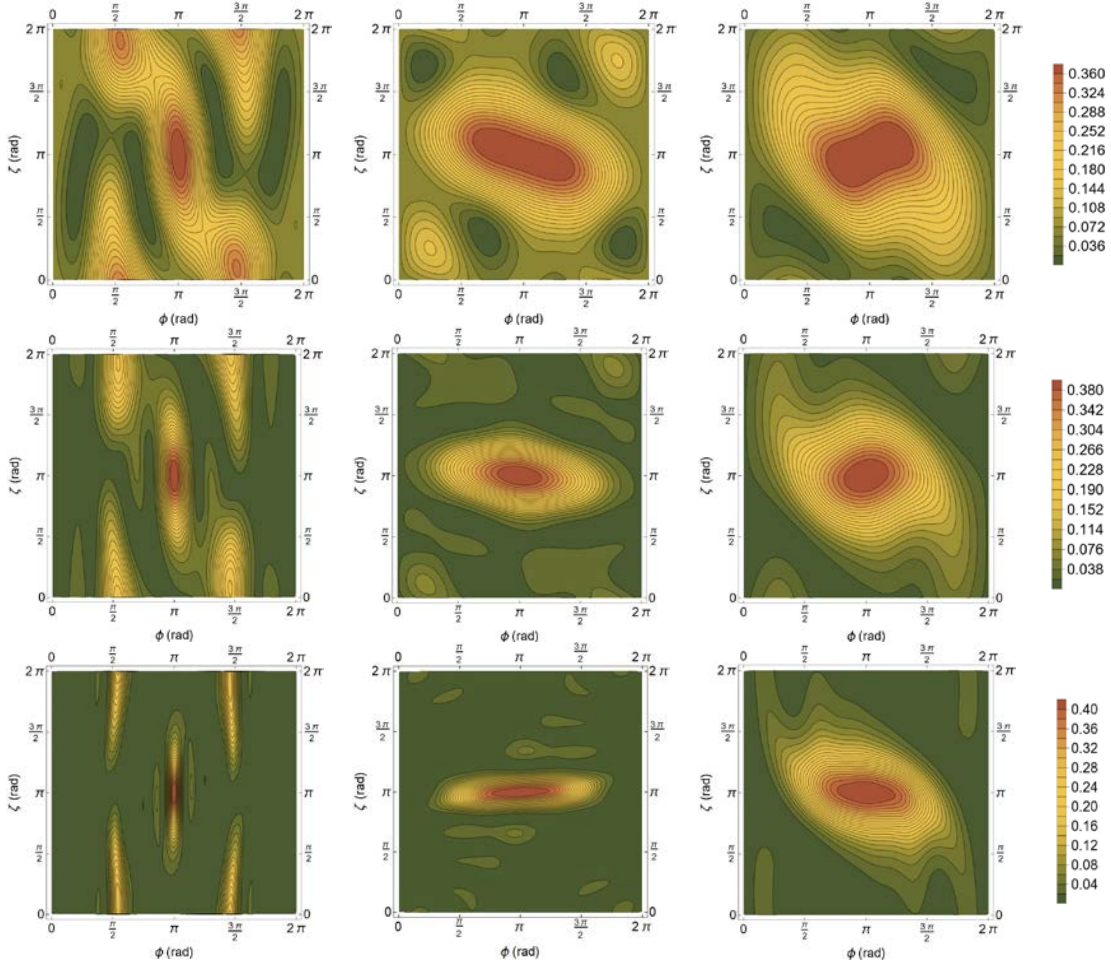

*Fig. 7. Probability to find solution of QRWS on hypercube for different sequences of walk coins constructed by using two phases ϕ and ζ. The first, second, third and fourth columns show results of simulations for sequences A1, A2, A3 and the colour legend. The first, second and third rows corresponds to coin sizes 4,6 and 8.*

On *Fig. 7* is shown the probability to find solution of the QRWS depending on the values of the phases ϕ and ζ for different sequences of phases in the coins. Different rows correspond to different coin size – the first to $m = 4$, the second to $m = 6$ and the third to $m = 8$. The first, second and third columns show the results for sequences (37), (38), (39) correspondingly. The last column shows the legend for that row – the magnitude of the probability to find solution, represented by color mapping.

It can be seen that for some cross-sections (linear functional dependences between the phases) the algorithm has higher robustness, for other – lower robustness. Both the optimal and the worst functions depend on both the register size and the chosen sequence.

On *Fig. 8* are shown as an example the results from simulations of $p(\phi(\omega), \zeta(\omega))$. Functions $\phi(\omega) = \omega \cos(\theta = const)$ and $\zeta(\omega) = \omega \sin(\theta = const)$ are chosen in such way that they are equivalent to linear dependence between phases that pass-through the point ($\phi = \pi, \zeta = \pi$). Here we will show results for two of them – the first one that gives the highest robustness (shown in red) and the second one with the lowest robustness (shown in green). Their corresponding Hill fits are shown with golden and teal. On the first row are shown results for coin size 4, on the second – for coin size 6 and on the third – for coin size 8. Simulations for dependences A1, A2 and A3 are shown on the first, second and third columns respectively.

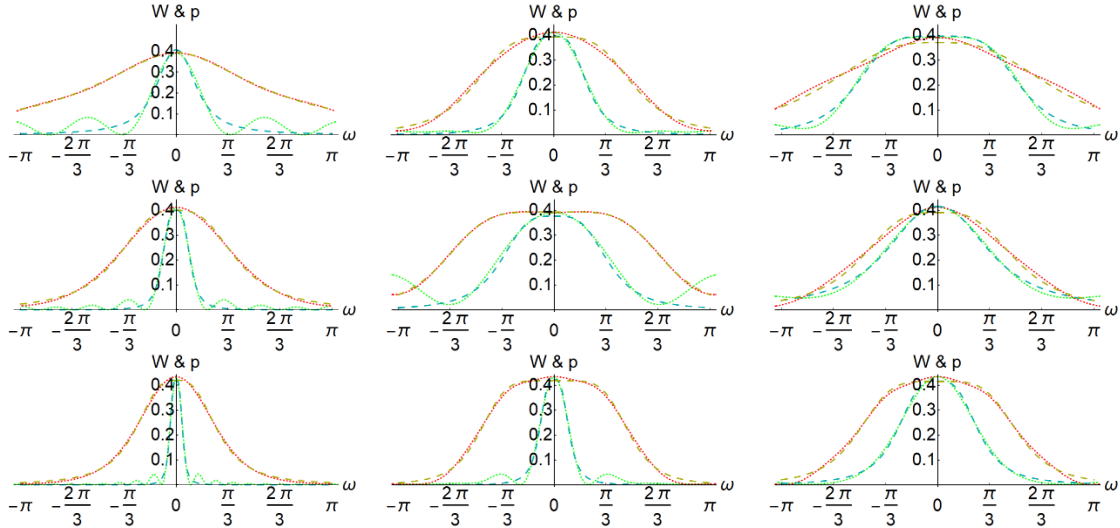

*Fig. 8. Results for $p(\phi(\omega), \zeta(\omega))$ and their Hill fit $W(\phi(\omega), \zeta(\omega))$ for the sequences: A1 (left column), A2 (central column) and A3 (right column). The first, second and third rows show simulations for coin size 4,6 and 8. The red line corresponds to the case of $p(\phi(\omega), \zeta(\omega))$ that has Hill fit (shown in golden) with maximal value of k. Similarly, the green line shows the case when $p$ has Hill fit (shown in teal) with minimal value of k*

We see that both the most and the least robust curves are well fitted by the Hill function. Results of the simulations and their fits with Hill function give approximately the minimal and maximal robustness obtained for different sequences and functional dependences. All other linear functional dependences for given sequence lie in those borders. This allows us to assess the robustness of the QRWS algorithm not only to errors in the phases ω, but to errors in the very dependence between the parameters $\theta$ too. Those analysis require knowing two of the parameters of Hill's function fit, namely k that estimates the robustness and b that gives estimation of the probability. By comparing those parameters for different coin sizes, we also can analyze how the robustness changes with increasing the coin size and make prognosis for even larger coins. Other important parameter is the value of ω that gives the highest robustness $ω_{max}$. We will use it to analyze how $Θ$ changes with increasing the coin size in order to achieve the highest robustness $p(ω_{max}, Θ = const)$.

On *Fig. 9* are shown the values of parameter b (top left picture), k (top right picture), n (bottom left picture) and $ω_{max}$ (bottom right picture) for different sequences and coin sizes between 4 and 9. For each sequence is used different colour and marker. Sequences A1, A2 and A3 are shown with blue (3-pointed star), purple (4-pointed star) and red (5-pointed star). The solid line represents the results of the fit with maximal value of k, and the dashed - the results of the fit with minimal value of k.

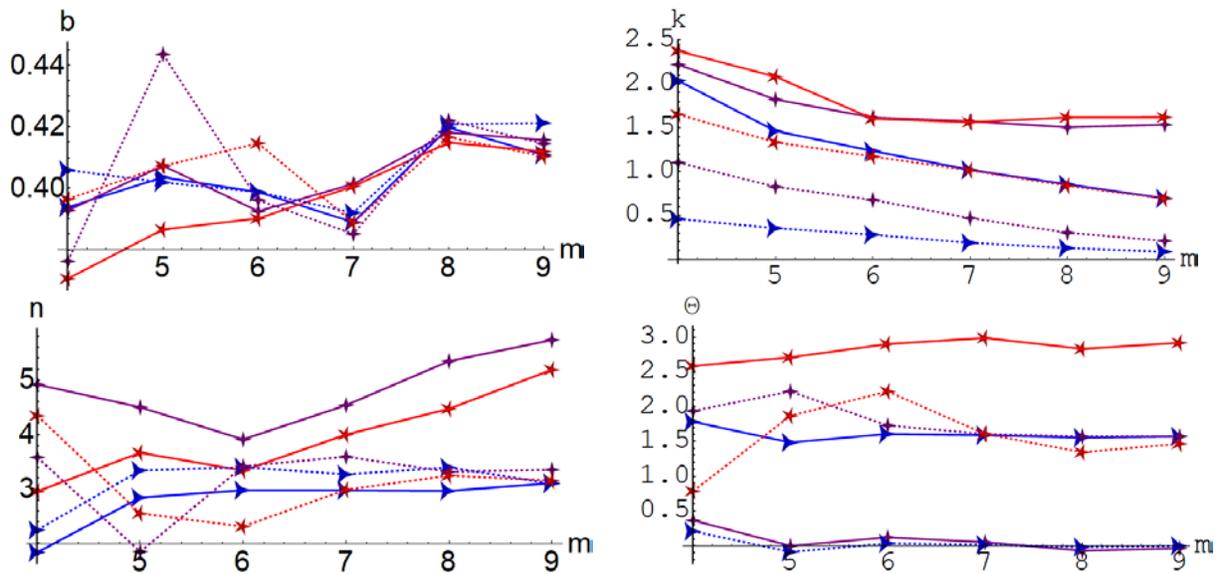

*Fig. 9. Values of the parameters of the Hill fit (b, k and n) for sequences A1, A2 and A3 in cases of highest robustness (shown with solid line) and worst robustness (shown with dotted line). Different colour and symbol represent different sequence – the blue colour (and three-point star), the purple colour (and four-point star) and the red colour and (five-point star) show sequences A1, A2 and A3 correspondingly.*

By using a limited number of values of the parameters b, k and n obtained from our simulations, we can make some observations and predictions:

The parameter k quantifies the robustness of the sequence. The best sequence is the one that has high robustness (large k) even for the functional dependence with the lowest value of k. We see that the worst sequence is the one when the generalized Householder angle ϕ remains constant. The best is when both angles ϕ and ζ change their sign each iteration. In the case of the best functional dependences for sequences A2 and A3, the parameter k goes to the same fixed value as the coin size increases. We can see that the worst functional dependence for the best sequence is as good as the best functional dependence for the worst sequence. In both latter cases the parameter k decreases with increasing the coin size.

The parameter b represents the average height of the plateau of the Hill function. This parameter for small coin sizes (4, 5 and 6) vary a lot, however for large coin sizes b has approximately the same value (around $b = 0.415$) for all sequences and coin sizes (7, 8 and 9).

The slope of the curve represented by the parameter n, is higher for sequences A2 and A3 and the best functional dependences between phases. For the worst $\Theta$, all sequences have the same value n.

The best angle $\Theta$ for small coin sizes vary. However, with increasing the coin size it goes to a fixed value. We extrapolate that each sequence goes to a different fixed value A1 to $\pi/2$, A2 to 0 and A3 to $\pi$. The angle of the worst functional dependence is perpendicular to the one with the best for all three sequences.

In this work, the focus of our research is the robustness, so from now on we will focus only on it. Hill fit can be used to compare how the robustness changes with increasing the register size for different sequences. We will use the following function to fit the parameter k in order to make extrapolations for it, in case of larger coin register sizes in both - the best and the worst cases:

$$k_{Sequence}^{Best\ or\ Worst}(m) = k_1 e^{-mk_2} + k_3 \qquad (40)$$

Table 3 shows the results of fitting the parameter k for different sequences (shown in second column) and functional dependences (shown in third column). The parameters of the fit are shown in the fourth, fifth and sixth columns. The standard deviation used to evaluate the goodness of the fit is shown in 7-th column.

| № | Sequence | $Best\ or\ Worst$ | $k_1$ | $k_2$ | $k_3$ | $\sigma$ |
|---|---|---|---|---|---|---|
| 1 | A1 | $Best$ | 6.93968 | 0.382892 | 0.512436 | 0.0557685 |
| 2 |    | $Worst$ | 1.30841 | 0.174113 | -0.188854 | 0.0089503 |
| 3 | A2 | $Best$ | 22.1818 | 0.858386 | 1.50822 | 0.0232775 |
| 4 |    | $Worst$ | 3.09085 | 0.168702 | -0.475159 | 0.02708 |
| 5 | A3 | $Best$ | 18.3219 | 0.763294 | 1.54483 | 0.12208 |
| 6 |    | $Worst$ | 3.18826 | 0.168853 | 0.0108582 | 0.0341235 |

*Table 3. Fit of the parameter k for sequences A1, A2 and A3 for the best and the worst angle $\Theta$. All sequences are fit by using the function shown in Eq.(40). The second column corresponds to the sequence. The third column shows if it the best functional dependence between the phases or its worst. The fourth, fifth and sixth columns give the parameters of the fit. The last column shows the standard deviation.*

The value of $k_3$ shows, that for sequence A3 in case of both best and worst angle, this modification gives a good result. For the other two sequences the modification gives good result only for the optimal functional dependence. The function used for fitting allows us to predict the parameter values for each of the sequence and functional dependencies studied. The value of Eq.(40) when $m \to \infty$ goes to the value of $k_3$.

In the next section we will study 3 more sequences and then we will compare the results for all sequences studied in this work.

## 9. Sequences when each half of iterations has same sign

In this section will study sequences that have a certain sign of the phase in the first half of the iterations, and in the second half – the sign is reversed. More specific sequences explored here are:

$$\{\phi_j, \zeta_j\} = \{\phi, (-1)^{\|j/k_{iter}\|}\zeta\} \quad denoted\ as\ H1 \tag{41}$$

$$\{\phi_j, \zeta_j\} = \{(-1)^{\|j/k_{iter}\|}\phi, \zeta\} \quad denoted\ as\ H2 \tag{42}$$

$$\{\phi_j, \zeta_j\} = \{(-1)^{\|j/k_{iter}\|}\phi, (-1)^{\|j/k_{iter}\|}\zeta\} \quad denoted\ as\ H3 \tag{43}$$

where $\|j/k_{iter}\|$ means that $j/k_{iter}$ is rounded.

The probability to find solution when each half iteration of QRWS has the same sign, depending on the values of the phases ф and ζ for different sequences, is shown on *Fig. 10*. Again, the rows correspond to different coin size – on the first, second and third rows are shown the simulations for $m = 4$, $m = 6$ and to $m = 8$ correspondingly. Results for the sequences of phases according to equations (41), (42) and (43) are shown on the first, second and third columns respectively. In the fourth column is the legend shoving the magnitude of probability to find solution in color mapping.

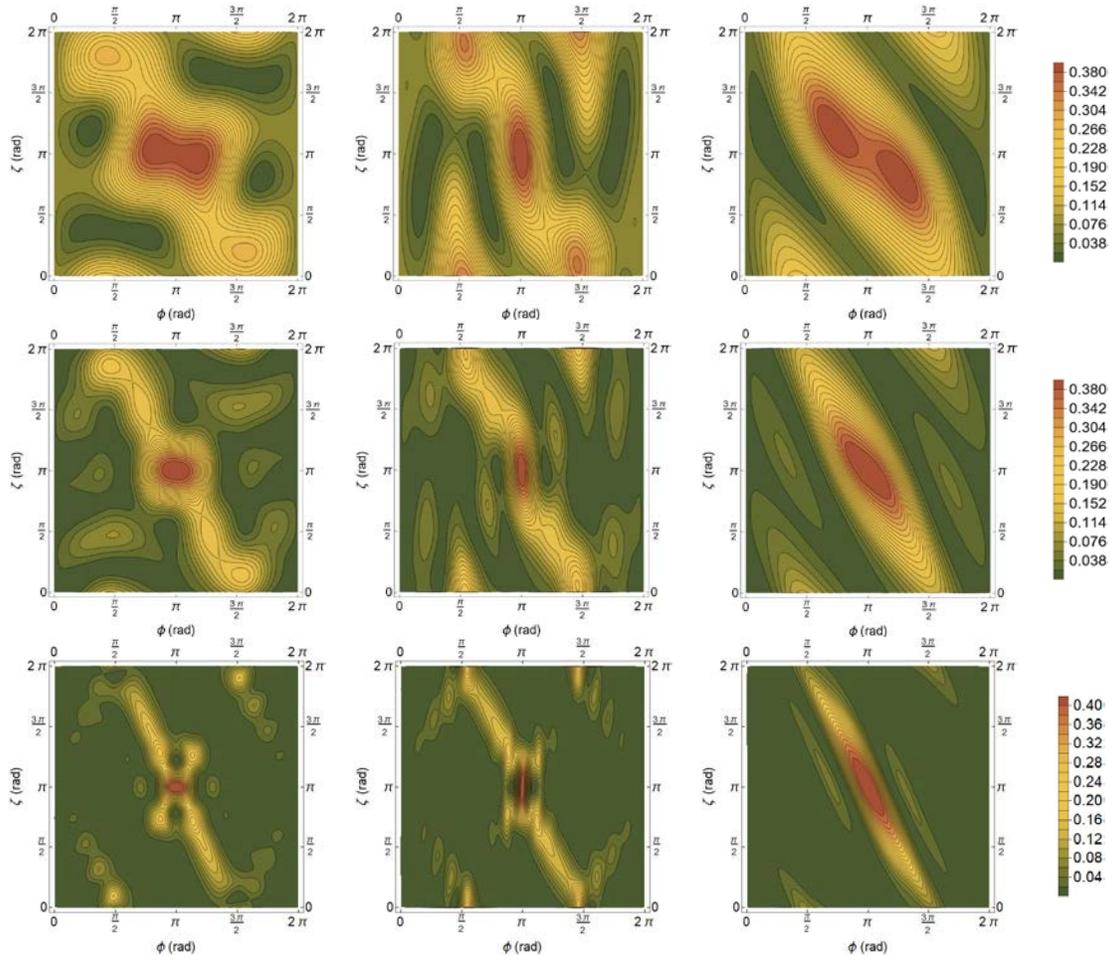

*Fig. 10. Probability to find solution of QRWS algorithm on Hypercube depending on the phases ϕ and ζ in case of different sequences. The horizontal axis on each picture shows the phase φ and the vertical one - the phase ζ. On the first, second, and third columns are given pictures for H1, H2 and H3 sequences. The fourth column shows the correspondence between the color and the probability. In the first, second and third rows are shown results for coin size 4, 6 and 8.*

Similarly, to the previous section, we simulate $p(\phi(\omega), \zeta(\omega))$ when $\phi(\omega)$ and $\zeta(\omega)$ are equivalent to a function that passes through the point ($\phi = \pi, \zeta = \pi$). On *Fig. 11* we show the results for simulations in case of coin size 4 (first row), 6 (second row) and 8 (third row). The cases of highest robustness are shown in red dots and the lowest are shown in green dots, while their corresponding fits by Hill function are shown with golden and teal colours. The first second and third columns show the results for H1, H2 and H3 respectively.

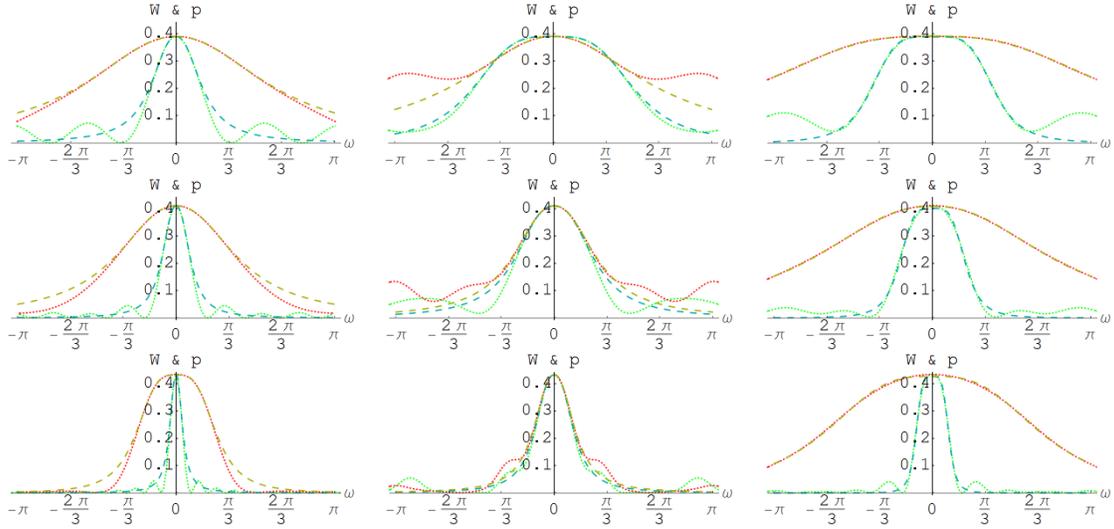

*Fig. 11. Probability to find solution of QRWS on Hypercube when both phases (ϕ and ζ) depend on the parameter ω according to Eq. (18) for different sequences of phases. The left column corresponds to sequence H1, the central to H2 and the right one - to H3. The first, second and third rows show the results from simulations for coin sizes 4, 6 and 8 respectively. The red dotted line represents the success probability of QRWS for the optimal phase of the sequences and the green dotted line - the worst phase for the sequences. Their corresponding Hill fits are shown with golden and teal colors.*

Hill function fits well the results for QRWS success probability for coin sizes between 4 and 9. This means that we can use the parameters of the fit in order to compare the advances of the different sequences and to investigate how the robustness, average height and slope change with increasing the coin size.

*Fig. 12* shows values of the Hill function's parameters used to fit sequences H1, H2 and H3 for coin sizes between 4 and 9. The top left and right pictures show result for the parameters b and k, while the bottom left and right pictures present the results for n and $\Theta_{max}$ respectively. Sequences H1, H2 and H3 are depicted with blue triangle, purple square and red pentagon correspondingly. Results of the simulations are represented with solid lines in case of maximal robustness and in case of minimal robustness - with dashed lines.

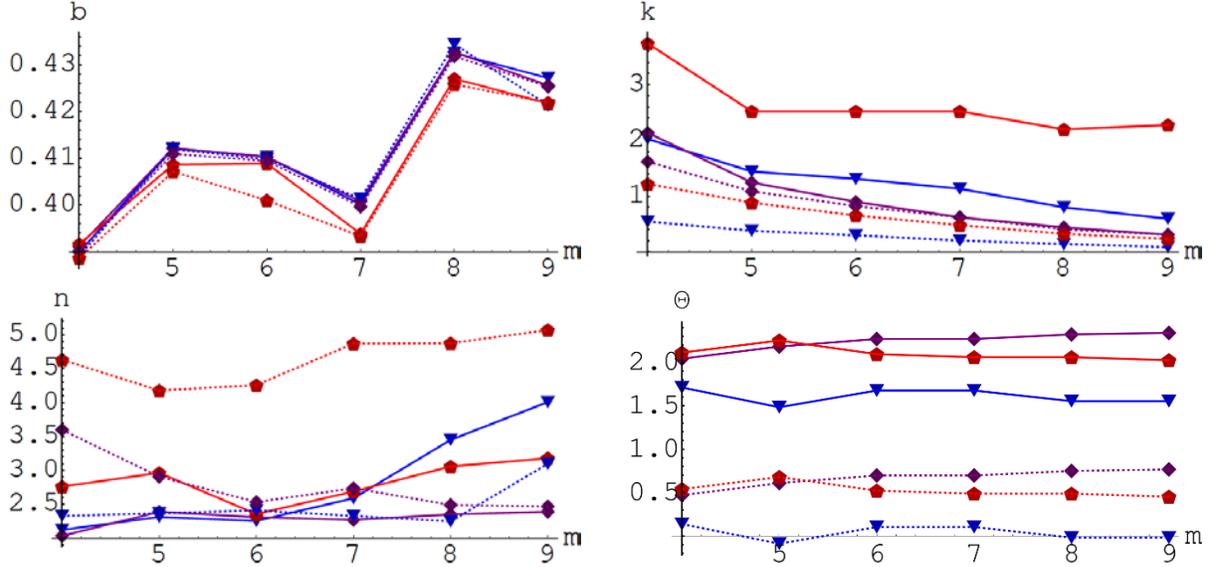

*Fig. 12. Hill fit parameter's values in case of highest and lowest robustness for sequences H1, H2 and H3. The parameters b and k are shown on the top left and right pictures, the parameter n on the bottom left and the angle Θ on the bottom right. The values for the highest robustness are shown with solid line and dashed line shows the values with the lowest robustness. The numerical results for sequences H1, H2 and H3 are given with a triangle, square and pentagon respectively.*

Similarly, to the case of A sequences, we can use the limited number of points in order to obtain some predictions.

The numerical results show that the average value of $b$ is also approximately the same for A and H modifications. The largest slope (correspond to n) is for the worst angle Θ and sequence H3.

We extrapolate that the angle Θ of each sequence, as $m \to \infty$, goes to a different fixed value H1 to $\pi/2$, H2 to $7\pi/9$ and H3 to $2\pi/3$. The angle of the worst functional dependence is perpendicular to the one with the best of all three sequences.

In case of optimal angle Θ, the parameter k is largest in case of H3, however in case of worst value of Θ, the sequence H3 has very low robustness. The sequence H2 has the same values of the robustness in both cases of optimal and worst value of Θ for coin sizes up to 9. However, in order to see if this behaviour preserves with increasing the coin size, we should make an extrapolation for larger coin size. Here again we will use the same function as in the previous section in order to fit the Hill function parameter k. This will allow us to easily compare both sequences (i.e. A1, A2, A3 and H1, H2, H3):

$$k_{Sequence}^{Best\ or\ Worst}(m) = k_1 e^{-mk_2} + k_3 \qquad (44)$$

The results of fitting the values of the parameter k for sequences H1, H2 and H3 for coin sizes between 4 and 9 are shown in *Table 4*. The sequence is in the second column. Whether these are the parameters for the most or least robust fit is shown in the third column. The fourth, fifth and six columns give the parameters of the fit. The goodness of the fit is evaluated by standard deviation and is positioned in the seventh column.

| № | Sequence | *Best or Worst* | $k_1$ | $k_2$ | $k_3$ | $\sigma$ |
|---|---|---|---|---|---|---|
| 1 | H1 | *Best* | 4.59679 | 0.177573 | -0.295823 | 0.116051 |
| 2 |    | *Worst* | 1.7612 | 0.264969 | -0.0702085 | 0.0114705 |
| 3 | H2 | *Best* | 17.0124 | 0.548523 | 0.22363 | 0.0573324 |
| 4 |    | *Worst* | 6.60375 | 0.364563 | 0.0686182 | 0.0394462 |
| 5 | H3 | *Best* | 190.244 | 19.1032 | 2.63443 | 0.725247 |
| 6 |    | *Worst* | 4.09298 | 0.288725 | 0.0706958 | 0.0101359 |

*Table 4. Fit of the parameter k by using the function Eq. (44), in case of sequences H1, H2 and H3. The fits are made in case of both optimal and worst values of the parameter* Θ. *In the second and third column are shown the sequence and whether* Θ *is optimal or worst. The fourth, fifth and sixth columns give the parameters of the fit. The standard deviation is shown in the eighth column.*

Both fits of H1 by *Eq.*(40) give unrealistic prognose for $m \to \infty$. This means that in this case the obtained fitting function should not be used for register size significantly exceeding the simulated values ($4 \leq m \leq 9$). However, in the cases of H2 and H3 the fits show that H3 is again much better than H2. We can also see that the bad dependence for H3 is approximately as good as the good dependence for H2.

In the next section we will compare the best A and H fits with the phase matching sequence in order to analyze their advantages and disadvantages.

### 10. Comparison of the modification's robustness

In this section we summarize the results and compare the different multi-phase matching sequences with the phase matching case. We will make the comparison in both cases of the maximal robustness of the sequence ($\Theta = \Theta_{MAX}$) and of the minimal robustness ($\Theta = \Theta_{MIN}$).

Our results show that in the case when the best dependence between the phases is implemented, the maximal robustness is achieved in phase matching modification (with prognosed value for large coin size), followed by all multi-phase matching sequences. The best sequence is H3 (when m is large), followed by sequences A3 (for large coin) and A1 (for large coin).

In the case of the worst dependence between the parameters, the highest robustness is achieved for sequence A3, followed by H3 and A2. In this case the phase matching modification (Section 6) has very low stability. This trend strengthens as the size of the coin increases.

Which is the best sequence depends on how well we can maintain the dependence between the phases ϕ and ζ. If we can preserve the exact relations, then the best QRWS modification is the phase matching one. In case when we can't maintain it, the sequence A3 is the best.

## 11. Conclusion

In this work we study QRWS algorithm with traversing coins constructed by using generalized Householder reflection and a phase multiplier, in the cases when the sign of the phases depends on the iteration number (called sequence of phases in the paper). We investigate numerically, six different sequences – three with alternatively changing the sign each iteration and three when first half of the iterations the sign is one and the second half has the opposite sign. By using semi empirical methods, we have obtained interpolations (for coin size up to 9) for the different characteristics, like the robustness against inaccuracies in the phases of the traversing coin and the average height in the interval where the function is robust. Next, we extrapolate the robustness for larger coin size. Our results show that by using multi-phase matching, the algorithm becomes more robust against deviations in the optimal functional dependencies between phases at the cost of robustness reduction in the case when this dependency is strictly preserved.

**Acknowledgments:**

This work was supported by the Bulgarian National Science Fund under contract KP-06-N58/5 / 19.11.2021.

**References**:

[1] N. Shenvi, J. Kempe, and K. B. Whaley, "Quantum random-walk search algorithm," *Phys. Rev. A*, vol. 67, no. 5, p. 052307, May 2003, doi: 10.1103/PhysRevA.67.052307.

[2] T. G. Wong and A. Ambainis, "Quantum search with multiple walk steps per oracle query," *Phys. Rev. A*, vol. 92, no. 2, p. 022338, Aug. 2015, doi: 10.1103/PhysRevA.92.022338.

[3] V. Potoček, A. Gábris, T. Kiss, and I. Jex, "Optimized quantum random-walk search algorithms on the hypercube," *Phys. Rev. A*, vol. 79, no. 1, p. 012325, Jan. 2009, doi: 10.1103/PhysRevA.79.012325.

[4] A. Ambainis, J. Kempe, and A. Rivosh, "Coins Make Quantum Walks Faster." arXiv, Feb. 16, 2004. doi: 10.48550/arXiv.quant-ph/0402107.

[5] D. Koch and M. Hillery, "Finding paths in tree graphs with a quantum walk," *Phys. Rev. A*, vol. 97, no. 1, p. 012308, Jan. 2018, doi: 10.1103/PhysRevA.97.012308.

[6] A. Patel and K. S. Raghunathan, "Search on a fractal lattice using a quantum random walk," *Phys. Rev. A*, vol. 86, no. 1, p. 012332, Jul. 2012, doi: 10.1103/PhysRevA.86.012332.

[7] J. K. Gamble, M. Friesen, D. Zhou, R. Joynt, and S. N. Coppersmith, "Two-particle quantum walks applied to the graph isomorphism problem," *Phys. Rev. A*, vol. 81, no. 5, p. 052313, May 2010, doi: 10.1103/PhysRevA.81.052313.

[8] A. Ambainis, A. M. Childs, B. W. Reichardt, R. Špalek, and S. Zhang, "Any AND-OR Formula of Size N Can Be Evaluated in Time $N^{1/2+o(1)}$ on a Quantum Computer," *SIAM J. Comput.*, vol. 39, no. 6, pp. 2513–2530, Jan. 2010, doi: 10.1137/080712167.

[9] F. Magniez, M. Santha, and M. Szegedy, "Quantum Algorithms for the Triangle Problem," *SIAM J. Comput.*, vol. 37, no. 2, pp. 413–424, Jan. 2007, doi: 10.1137/050643684.

[10] D. Kielpinski, C. Monroe, and D. J. Wineland, "Architecture for a large-scale ion-trap quantum computer," *Nature*, vol. 417, no. 6890, Art. no. 6890, Jun. 2002, doi: 10.1038/nature00784.
[11] P. Kok, W. J. Munro, K. Nemoto, T. C. Ralph, J. P. Dowling, and G. J. Milburn, "Linear optical quantum computing with photonic qubits," *Rev. Mod. Phys.*, vol. 79, no. 1, pp. 135–174, Jan. 2007, doi: 10.1103/RevModPhys.79.135.
[12] P. A. Ivanov and N. V. Vitanov, "Synthesis of arbitrary unitary transformations of collective states of trapped ions by quantum Householder reflections," *Phys. Rev. A*, vol. 77, no. 1, p. 012335, Jan. 2008, doi: 10.1103/PhysRevA.77.012335.
[13] E. S. Kyoseva, D. G. Angelakis, and L. C. Kwek, "A single-interaction step implementation of a quantum search in coupled micro-cavities," *EPL*, vol. 89, no. 2, p. 20005, Feb. 2010, doi: 10.1209/0295-5075/89/20005.
[14] P. A. Ivanov, E. S. Kyoseva, and N. V. Vitanov, "Engineering of arbitrary $\mathrm{U}(N)$ transformations by quantum Householder reflections," *Phys. Rev. A*, vol. 74, no. 2, p. 022323, Aug. 2006, doi: 10.1103/PhysRevA.74.022323.
[15] M. A. Nielsen and I. L. Chuang, *Quantum computation and quantum information*. Cambridge: Cambridge Univ. Press, 2007.
[16] N. V. Vitanov, "Synthesis of arbitrary SU(3) transformations of atomic qutrits," *Phys. Rev. A*, vol. 85, no. 3, p. 032331, Mar. 2012, doi: 10.1103/PhysRevA.85.032331.
[17] J. Kempe, "Quantum random walks - an introductory overview," *Contemporary Physics*, vol. 44, no. 4, pp. 307–327, Jul. 2003, doi: 10.1080/00107151031000110776.
[18] T. A. Brun, H. A. Carteret, and A. Ambainis, "Quantum walks driven by many coins," *Phys. Rev. A*, vol. 67, no. 5, p. 052317, May 2003, doi: 10.1103/PhysRevA.67.052317.
[19] L. Innocenti *et al.*, "Quantum state engineering using one-dimensional discrete-time quantum walks," *Phys. Rev. A*, vol. 96, no. 6, p. 062326, Dec. 2017, doi: 10.1103/PhysRevA.96.062326.
[20] C. Vlachou, J. Rodrigues, P. Mateus, N. Paunković, and A. Souto, "Quantum walk public-key cryptographic system," *Int. J. Quantum Inform.*, vol. 13, no. 07, p. 1550050, Oct. 2015, doi: 10.1142/S0219749915500501.
[21] C. Vlachou, W. Krawec, P. Mateus, N. Paunković, and A. Souto, "Quantum key distribution with quantum walks," *Quantum Inf Process*, vol. 17, no. 11, p. 288, Sep. 2018, doi: 10.1007/s11128-018-2055-y.
[22] H.-J. Li, J. Li, N. Xiang, Y. Zheng, Y.-G. Yang, and M. Naseri, "A new kind of universal and flexible quantum information splitting scheme with multi-coin quantum walks," *Quantum Inf Process*, vol. 18, no. 10, p. 316, Aug. 2019, doi: 10.1007/s11128-019-2422-3.
[23] D. Nelson and M. Cox, "Lehninger Principles of Biochemistry, 8th Edition | Macmillan Learning CA." Accessed: May 21, 2023. [Online]. Available: https://www.macmillanlearning.com/college/ca/product/Lehninger-Principles-of-Biochemistry/p/1319228003
[24] U. Alon, *An Introduction to Systems Biology: Design Principles of Biological Circuits*. Accessed: Dec. 09, 2022. [Online]. Available: https://www.routledge.com/An-Introduction-to-Systems-Biology-Design-Principles-of-Biological-Circuits/Alon/p/book/9781439837177
[25] H. Tonchev and P. Danev, "Studying the robustness of quantum random walk search on Hypercube against phase errors in the traversing coin by semi-empirical methods," in *Proceedings of 11th International Conference of the Balkan Physical Union — PoS(BPU11)*, vol. 427, SISSA Medialab, 2023, p. 175. doi: 10.22323/1.427.0175.

[26] M. Piatek and A. R. Pietrykowski, "Classical irregular blocks, Hill's equation and PT-symmetric periodic complex potentials," *J. High Energ. Phys.*, vol. 2016, no. 7, p. 131, Jul. 2016, doi: 10.1007/JHEP07(2016)131.

[27] J. P. Gaebler et al., "High-Fidelity Universal Gate Set for ${^{9}\mathrm{Be}}^{+}$ Ion Qubits," *Phys. Rev. Lett.*, vol. 117, no. 6, p. 060505, Aug. 2016, doi: 10.1103/PhysRevLett.117.060505.

[28] M. Brownnutt, M. Kumph, P. Rabl, and R. Blatt, "Ion-trap measurements of electric-field noise near surfaces," *Rev. Mod. Phys.*, vol. 87, no. 4, pp. 1419–1482, Dec. 2015, doi: 10.1103/RevModPhys.87.1419.

[29] W. Wei et al., "Measurement and suppression of magnetic field noise of trapped ion qubit," *J. Phys. B: At. Mol. Opt. Phys.*, vol. 55, no. 7, p. 075001, Apr. 2022, doi: 10.1088/1361-6455/ac5e7d.

[30] "Noise effect on Grover algorithm | The European Physical Journal D." Accessed: Apr. 03, 2024. [Online]. Available: https://link.springer.com/article/10.1140/epjd/e2007-00295-1

[31] A. Bassi and D.-A. Deckert, "Noise gates for decoherent quantum circuits," *Phys. Rev. A*, vol. 77, no. 3, p. 032323, Mar. 2008, doi: 10.1103/PhysRevA.77.032323.

[32] V. Kendon, "Decoherence in quantum walks – a review," *Mathematical Structures in Computer Science*, vol. 17, no. 6, pp. 1169–1220, Dec. 2007, doi: 10.1017/S0960129507006354.

[33] C. M. Chandrashekar, R. Srikanth, and S. Banerjee, "Symmetries and noise in quantum walk," *Phys. Rev. A*, vol. 76, no. 2, p. 022316, Aug. 2007, doi: 10.1103/PhysRevA.76.022316.

[34] K. Manouchehri and J. B. Wang, "Quantum walks in an array of quantum dots," *J. Phys. A: Math. Theor.*, vol. 41, no. 6, p. 065304, Jan. 2008, doi: 10.1088/1751-8113/41/6/065304.

[35] C.-F. Chiang and C.-Y. Hsieh, "Noise Characterization: Keeping Reduction Based Perturbed Quantum Walk Search Optimal," *EPJ Web Conf.*, vol. 198, p. 00001, 2019, doi: 10.1051/epjconf/201919800001.

[36] Y.-C. Zhang, W.-S. Bao, X. Wang, and X.-Q. Fu, "Effects of systematic phase errors on optimized quantum random-walk search algorithm*," *Chinese Phys. B*, vol. 24, no. 6, p. 060304, Apr. 2015, doi: 10.1088/1674-1056/24/6/060304.

[37] H. Tonchev and P. Danev, "Optimizing the walk coin in the quantum random walk search algorithm," *Int. J. Quantum Inform.*, vol. 21, no. 06, p. 2350030, Sep. 2023, doi: 10.1142/S0219749923500302.

[38] H. Tonchev and P. Danev, "A Machine Learning Study of High Robustness Quantum Walk Search Algorithm with Qudit Householder Coins," *Algorithms*, vol. 16, no. 3, Art. no. 3, Mar. 2023, doi: 10.3390/a16030150.

[39] H. Tonchev and P. Danev, "Robustness of different modifications of Grovers algorithm based on generalized Householder reflections with different phases." arXiv, Jan. 07, 2024. doi: 10.48550/arXiv.2401.03602.

[40] F. M. Toyama, W. van Dijk, Y. Nogami, M. Tabuchi, and Y. Kimura, "Multiphase matching in the Grover algorithm," *Phys. Rev. A*, vol. 77, no. 4, p. 042324, Apr. 2008, doi: 10.1103/PhysRevA.77.042324.

[41] H. Tonchev and P. Danev, "Reducing number of gates in quantum random walk search algorithm via modification of coin operators," *Results in Physics*, vol. 46, p. 106327, Mar. 2023, doi: 10.1016/j.rinp.2023.106327.

[42]    H. Tonchev and P. Danev, "Robustness of Quantum Random Walk Search Algorithm in Hypercube when only first or both first and second neighbors are measured." arXiv, May 24, 2023. doi: 10.48550/arXiv.2305.15073.